\documentstyle[emulateapj,apjfonts,epsfig]{article}
\newcommand{\Ne}{$^{22}{\rm Ne} \ $}
\begin{document}

\title{Gravitational Settling of $^{22}$Ne in Liquid White Dwarf Interiors}

\author{Lars Bildsten} 
\affil{Institute for Theoretical Physics and Department of Physics\\
Kohn Hall, University of California, Santa Barbara, CA 93106 \\
bildsten@itp.ucsb.edu\\}

\and

\author{David M. Hall} 
\affil{Department of Physics, Broida Hall, 
University of California, Santa Barbara, CA 93106 \\
halldm@physics.ucsb.edu\\}

\centerline{ To Appear in Astrophysical Journal Letters} 

\begin{abstract}

The nuclear reactions that occur in the stellar progenitors of white
dwarfs (WDs) lead to an internal composition of $^{12}{\rm C}$,
$^{16}{\rm O}$, and a ``contaminant'' nucleus, \Ne. The \Ne is
produced by helium captures on $^{14}{\rm N}$ left from hydrogen
burning via the CNO cycle.  By virtue of its two excess neutrons
(relative to the predominant $A=2Z$ nuclei), a downward force of
$\approx 2 m_p g$ is exerted on \Ne in the WD interior. This biases
its diffusive equilibrium, forcing \Ne to settle towards the center of
the WD. We discuss the physics of the gravitational settling when the
WD is in the liquid state and the luminosity generated by it. This
modifies the cooling of WD's with masses in excess of $M_\odot$. The
current uncertainties in the microphysics even allow for solutions
where a $1.2M_\odot$ WD remains mostly liquid for a few Gyrs due to
the internal heating from \Ne sedimentation. This highlights the need
for an accurate calculation of the inter-diffusion coefficient,
especially in the quantum liquid regime relevant for high mass WD's.
There is also time in old, liquid WD's (such as those found in
cataclysmic variables and possibly in accreting Type Ia progenitors)
for partial settling.

\end{abstract}



\section{Introduction}\label{sec:Intro}

 The slowest step in the CNO cycle is the proton
capture on $^{14}{\rm N}$. This results in all CNO catalysts piling up
at $^{14}{\rm N}$ when H burning is completed during the main
sequence. During Helium burning, the reactions $^{14} {\rm N}(\alpha,
\gamma) ^{18}{\rm F} (\beta^{+})^{18}{\rm O}(\alpha, \gamma)^{22}{\rm
Ne}$ ensue and convert the $^{14}{\rm N}$ into \Ne, resulting in a
mass fraction of \Ne, $X_{22}\approx Z_{CNO}\approx 0.02$ for
Population I.  The possibility that this nucleus plays an important
role in the energetics of cooling WD's was first noted by Isern et
al. (1991). They discussed the possibility of gravitational energy
release if \Ne phase separates at crystallization. This was followed
by more detailed studies (Xu \& Van Horn 1992; Ogata et. al. 1993;
Segretain et al. 1994; Segretain 1996) which differed in their
conclusions about the final state once crystallization was complete;
ranging from all the \Ne in the stellar center to most of it having an
unchanged profile. 

What about \Ne sedimentation in the liquid state?  Ogata et al. (1993)
showed that demixing in the liquid state never occurs for the C/O/\Ne
liquid (namely it will remain in the fully mixed state when the
temperature is safely above melting).  This allows us to focus on
the gravitational settling of individual
\Ne nuclei in the liquid parts of the white dwarf (Bravo et. al. 1992). 
This happens prior to the onset of cystallization in a 
cooling WD and during the lifetime of a WD that remains in the liquid
state due to heating from accretion.

Degenerate electrons dominate the pressure in the WD interior, so
that the upward pointing electric field at mass coordinate $m(r)$ is
$eE\approx 2m_pg$, where $g=Gm(r)/r^2$ and $m_p$ is the proton mass.
The downward force on a \Ne nucleus is $F=22m_pg-10eE=2m_pg$, giving a
potential energy drop $U(r)=\int_0^r 2m_pg dr$ between radius $r$ and
the center. Across the whole star, this
amounts to 285 keV
for $M=0.6 M_\odot$ and 1600 keV for $M=1.2M_\odot$.  It is the excess
neutrons of the \Ne nucleus (relative to $A=2Z$) that make it special
(the same is true of the less abundant $^{56}{\rm
Fe}$). Differentiating forces on the carbon and oxygen nuclei only
arise from the slight differences in masses from $Am_p$ due to the
nuclear binding, which is orders of magnitude smaller than the force
on \Ne.

 The potential energy (the energy that would be released if all the
\Ne sinks to the center) of a uniformly mixed model ($X_{22}=0.02$ is
constant) is $1.4\times 10^{47}, 3.2\times 10^{47}, 6.8\times 10^{47}$
and $1.5\times 10^{48} \ {\rm erg}$ for masses of $0.6, 0.8, 1.0, \ \&
1.2 M_\odot$. For 0.6 and $1.0M_\odot$, this corresponds to $5.8\times
10^{15} \& \ 1.7\times 10^{16} {\rm erg \ gr^{-1}}$ of possible energy
release from \Ne settling. At a temperature of $10^{7}$ K, the
specific heat per oxygen ion is $7.7\times 10^{13} {\rm erg \
gr^{-1}}$, so that the energy stored in \Ne is comparable to the
thermal energy content of a cooling WD at late times.

 In \S 2, we estimate the inter-diffusion
coefficient for a trace \Ne nucleus in the liquid C/O interior. The
settling timescale for \Ne is $\sim 10$ Gyr, making it critical for a
more accurate calculation of the inter-diffusion coefficient. The
settling time decreases for more massive WDs.  In \S 3, we speculate
on the modifications the \Ne settling is likely to have on accreting
WD's and maybe Type Ia supernovae. We calculate the
luminosity from the gravitational energy release of sinking \Ne in \S
4. {\it This affects the cooling of WD's prior to crystallization and can
possibly heat them to a level hot enough to maintain them in the
liquid state for a Hubble time.} 

\section{Diffusion and Settling of the \Ne}

We presume that the WD interior consists of a single
element (either $^{12}{\rm C}$ or $^{16}{\rm O}$) of mass $Am_p$ and
charge $Ze$, and that \Ne is a trace component at density
$n_{\rm Ne}=X_{22}\rho/22 m_p$. Our WD is constructed purely from the
degenerate electron equation of state. For the properties we need for
the \Ne diffusion problem, it is a fine approximation to neglect the
ionic contribution to the pressure. 

  Relative diffusion of charged species in a correlated liquid is
substantially different than the diffusion problem in the WD
atmosphere. In those less dense regions, traditional Coulomb cross
sections and ideal gas equations of state dominate (e.g. Fontaine \&
Michaud 1979; Vauclair \& Vauclair 1982). As discussed by Paquette et
al. (1986), there is substantial uncertainty in the diffusion
coefficients in the regions of the WD where the ion Coulomb coupling
becomes strong. For a classical one component plasma (OCP) with ion
separation, $a$, defined by $a^3=3/4\pi n_i$, where
$n_i=\rho/A m_p$, the importance of Coulomb physics for the ions is
measured by
\begin{equation}
\label{eq:gamma}
\Gamma\equiv {(Ze)^2\over a kT} =57.7\rho_6^{1/3}
\left(10^7 \ {\rm K}\over T\right) 
\left(Z\over
8\right)^2\left(16\over A\right)^{1/3},
\end{equation}
where $\rho_6=\rho/10^6 \ {\rm gr \ cm^{-3}}$. Crystallization occurs
when $\Gamma$ exceeds 173 (see Farouki \& Hamaguchi 1993 and
references therein).  We only consider settling of \Ne in the liquid
state, or $1<\Gamma<173$, the presumption being that the dramatic
increase in viscosity expected in the solid state will prohibit
further gravitational sedimentation.

  The inter-diffusion coefficient, $D$, of \Ne in a C/O mixture with
$\Gamma\sim 1-100$ has not been specifically calculated and so we
construct here our best estimate. One way to
estimate $D$ is to use the Stokes-Einstein relation for a particle of
radius $a_p$ (in this case \Ne) undergoing Brownian motion in a fluid
of viscosity $\eta$, which gives $D=kT/4\pi a_p \eta$ when the fluid
is allowed to ``slip'' at the particle/fluid interface. This estimate works
well in other liquids at small scales, where it is accurate at 
atomic dimensions (Hansen \& McDonald 1986). An alternate route
is to use the self-diffusion coefficient in the OCP
\begin{equation}
\label{eq:diff}
D\approx 3 \omega_p a^2 \Gamma^{-4/3},
\end{equation} 
calculated by Hansen, McDonald and Pollock (1975), where
$\omega_p^2=4\pi n_i (Ze)^2/Am_p$ is the ion plasma frequency.  How
well does this agree with the Stokes-Einstein estimate?  Using the 
numerically derived viscosity of the OCP, $\eta\approx 0.1\rho
\omega_p a^2(\Gamma/10)^{0.3}$ (Donko \& Nyiri 2000 and references
therein) and setting $a=a_p$, the derived $D=kT/4\pi a \eta$ agrees with
equation (\ref{eq:diff}) to about 20\% for all $\Gamma>10$. 

We also compared equation (\ref{eq:diff}) to Fontaine's (1987) rough
fitting formula for the interdiffusion coefficients and found
agreement to $\approx 30$\%. So, for now, we will use equation
(\ref{eq:diff}) for our estimate of $D$, keeping in mind that there
has yet to be an appropriate microphysical calculation. We will show
later that there is astrophysical motivation for a more
accurate calculation of the inter-diffusion
coefficients, especially in the quantum liquid regime that is relevant
for the massive WD's. For these WDs, the parameter $\hbar \omega_p/kT$
exceeds unity long before crystallization sets in (Chabrier, Ashcroft
\& DeWitt 1992; Chabrier 1993) and might well modify the diffusion
coefficient by order unity. 
 
 Since $kT\ll U(r)$ over most of the stellar radius (equivalent to
saying that the \Ne scale height is much less than the WD radius),
most of the drift of the \Ne is dominated by ``falling'' at the local
speed set by the diffusion coefficient $V\approx {2m_pgD/kT}$, which
with equation (\ref{eq:diff}) gives
\begin{equation}
V={18m_pg\over Ze \Gamma^{1/3} (4\pi \rho)^{1/2}}. 
\end{equation}
In mass coordinates, $m(r)$, the 
time it takes to fall from $m_1$ to $m_2$ is 
\begin{equation}
\label{eq:delt}
\Delta t=
\int_{m_1}^{m_2}{Ze\Gamma^{1/3}\over 18 m_p G
(4\pi\rho)^{1/2}}{dm(r)\over m(r)}.
\end{equation}
For a constant density star, this becomes 
$\Delta t\approx t_{s}\int {\rm d}\ln m(r)$,
where 
\begin{equation}
t_s\equiv {\Gamma^{1/3} Z\over 18}\left(e^2\over G m_p^2\right)^{1/2}
\left(1\over 4\pi G \rho\right)^{1/2}=13\  {\rm
Gyr}{Z\Gamma^{1/3}\over  6\rho_6^{1/2}}.
\end{equation} 
Using the central densities as an estimate, a pure oxygen $0.6
M_\odot$ WD at $T=10^8 \ {\rm K}$ (this temperature is
appropriate for rapidly accreting WD's) has $t_s\approx 19.6\ {\rm
Gyr}$, so that complete sedimentation is not likely for the most
common $0.6M_\odot$ WD's.  However, the strong density dependence
means that this settling time decreases for massive WD's. For example,
a pure oxygen $1.2(1.3)M_\odot$ WD at $10^8 {\rm K}$ has $t_s\approx
4.8 (3){\rm Gyr}$, in rough agreement with the previous simple
estimates of Bravo et al. (1992).

  Figure 1 shows the time it takes for a \Ne nucleus that starts at
the surface ($m(r)=M$) to fall to the location, $m(r)$. This is an
integration of equation (\ref{eq:delt}) with a model WD. The solid
(dashed) lines are for pure oxygen (carbon) and bracket the C/O
WD's. We show two masses, 0.6 and 1.2 $M_\odot$, making it clear that
\Ne settles faster in a massive WD. 

\section{Accreting White Dwarfs and Type Ia Progenitors} 

  Unlike cooling WDs, the interiors of massive, accreting WD's are
maintained in the liquid state by the heating from accretion. For
slowly accreting ($\dot M\sim 10^{-10} M_\odot {\rm yr}^{-1}$) WD's in
cataclysmic variables below the period gap, the core temperatures are
in the range $T\approx (1-3)\times 10^7 \ {\rm K}$ (Nomoto 1982), so
that they remain liquid while accreting for a few Gyrs. A $1 M_\odot$ 
pure oxygen WD in such a setting will undergo neon settling in the
outer layers as 
$t_s\approx 12.6 \ {\rm Gyr}$ at $T=3\times 10^7 \ {\rm K}$.

  Though the debate rages regarding appropriate Type Ia progenitor
models, most favor rapidly accreting WDs that ignite carbon once the
mass is near the Chandrasekhar value. The accretion rates considered
are too rapid to allow \Ne settling during this phase of
accretion. However, prior evolution at lower accretion rates or an
extended period as a cooling WD will have allowed for some Neon
settling. Though unclear just how important this will be, the
effects of \Ne settling for the Type Ia ignition of a
near-Chandrasekhar mass WD are important to note. The most obvious one
is that \Ne settling will 
develop a gradient in the electron mean molecular weight. 
This would modify the
convective criterion in the ignited core and possibly affect the
evolution from the thermal runaway to a propagating flame front
(Garcia-Senz \& Woosley 1995).\footnote{Our present understanding of
the microphysics does not predict the development of a pure \Ne
core. However, if this were to occur, it is unlikely to modify the
ignition conditions, as the electron 
energy threshold for the reaction \Ne+$e^{-}\rightarrow ^{22}{\rm F} +
\nu_e$ is 11.36 MeV, which requires a density in excess of $2.1\times
10^{10} \ {\rm g \ cm^{-3}}$. Carbon ignites at densities about an
order of magnitude less (Nomoto, Thielemann \& Yokoi 1984), so that
the initial trigger of carbon burning is unchanged. However, a pure
\Ne core with a mass of $0.028 M_\odot$ would have a radius of $R_{\rm
Ne}\approx 190 \ {\rm km}$ at a density of $2\times 10^9 \ {\rm g \
cm^{-3}}$, forcing the C/O mixture to ignite in the shell covering the
\Ne core. }

The \Ne abundance in a Type Ia progenitor is also critical to the
overproduction of the neutron rich isotopes $^{54}$Fe and $^{58}$Ni
(Thielemann, Nomoto \& Yokoi 1986).  The neutron excess in the inner
parts of an exploding WD is fixed by electron captures. Modifying the
production of these elements at that location would require a
substantially enhanced \Ne abundance there (e.g. a pure \Ne core would
set the initial $Y_e\approx 0.46$). Outside of $(0.3-0.4)M_\odot$, the
production of $^{54}$Fe and $^{58}$Ni is fixed by the local \Ne
abundance (Iwamoto et al. 1999). Hence, the production of these
elements can be reduced if the \Ne sinks away from the outer regions
(Bravo et al. 1992).
\footnote{ H\"oflich et al. (1998, 2000) studied
the differences in Type Ia spectral evolution that are possible if the
$^{54}$Fe was absent in the outer layers. These can be large in the
UV, where the iron is a major opacity source.  Lentz et al. (2000)
varied the $^{54}$Fe abundance in the partially burned outer layer of
model W7 to see if the observations can accomodate a varying
value. Though they also found spectral differences in the UV, nothing
large enough to allow them to constrain the \Ne abundance.} Our
calculation suggests there is time for \Ne settling in the outer
liquid layers of a massive WD if accreting at a low rate. The final
answer to this puzzle awaits a time-dependent evolution on a
cosmological timescale of the diffusion equation with a realistic
starting abundance profile and accretion history.

 There is also a \Ne settling calculation relevant to classical novae.
Many lines of evidence (ejected masses and overly abundant heavy
elements) point to the ejected material in classical novae containing
matter dredged up from the underlying C/O WD (see Gehrz et al. 1998
and Starrfield 1999 for overviews).  Livio \& Truran (1994) noted that
the Neon abundances measured in many of these ejecta are consistent
with the $X_{22}\approx 0.02$ expected in a C/O WD. Hence, there is no
need to invoke the presence of an O/Ne/Mg WD for those
events. However, this will remain true only if the \Ne has not
gravitationally settled out of the outer C/O layers of the WD that are
about to be dredged up.

 We can assess this by finding how far the \Ne falls in the surface
C/O layers, where $g$ is constant. This is a simple case, where we
integrate equation (\ref{eq:delt}) for an isothermal layer. Using
pressure as a coordinate, we find that the time
it takes for \Ne to fall to a pressure $P$ starting from the surface is
\begin{equation}
\Delta t \approx 0.8 \ {\rm Gyr}\ \Gamma_{\rm P}^{1/3}
\left(Z\over 6\right)
\left(10^8 \ {\rm cm \ s^{-2}}\over g\right)^2
 \left(P\over 10^{20} \ {\rm erg \
cm^{-3}}\right)^{0.7}. 
\end{equation}
where $\Gamma_{\rm P}$ is the value of $\Gamma$ at pressure $P$ and 
exceeds one at the time of ignition. 

  Since the accreted material appears to dredge up a WD layer of mass
comparable to the accreted layer, we just need to compare the \Ne
settling time to the time it takes to accumulate the fuel, $t_a=4\pi
GM P/g^2 \dot M$. The ratio of these timescales gives
\begin{equation}
{\Delta t\over t_a}\approx 150 \Gamma_{\rm P}^{1/3} \left(M_\odot\over
M\right)\left(\dot M\over 10^{-9} \ {\rm M_\odot \ yr^{-1}}\right)
\left(2\times 10^{19} \ {\rm erg \ cm^{-3}}\over P\right)^{0.3},
\end{equation}
where we have written the pressure in units of the typical ignition
pressure (Livio 1994). Thus, for $\dot M> 10^{-11} M_\odot {\rm
yr^{-1}}$ (which applies to most classical novae, Livio 1994), the
underlying C/O material should have \Ne present at
$X_{22}\approx 0.02$, {\it allowing for neon enrichment on a C/O WD in those
classical novae that are constantly undergoing dredge-up.}

\section{Luminosities from Falling \Ne in Cooling White Dwarfs: Sedimentars} 

The rate of energy release from the falling \Ne is given by the 
integral of the power, $FV$, 
\begin{equation}
L_g\equiv \int F V n_{\rm Ne} 4\pi r^2 dr. 
\end{equation}
Since the \Ne settling time is longer than the time it takes to cool
to the onset of crystallization in the core (about 2(0.3) Gyr for a
$0.6 (1.2) M_\odot$ WD; Benvenuto \& Althaus 1999), we presume that
the abundances are nearly unchanged from the initial state. This
integral is evaluated with $X_{22}=0.02$ and the core temperature,
$T_c$, being constant.  For a pure oxygen WD at $T_c=10^8$ K, this
gives $L_g=1.1\times 10^{29}, 3.9\times 10^{29}, 1.2\times 10^{30} \&
4.6\times 10^{30}\ {\rm erg \ s^{-1}}$ for $M=0.6, 0.8, 1.0, \ \& \
1.2 M_{\odot}$.

For massive WD's, $L_g$ becomes comparable
to the cooling luminosity, $L$, for the same $T_c$. This raises the
important question of how much the settling luminosity can affect the
WD cooling curves and thus the inferred age of the galactic disk from
the faint end of the WD luminosity function (see Leggett, Ruiz and
Bergeron 1998 for a recent overview).

 We begin with the common $0.6 M_\odot$ WDs. The hatched region in the
top panel of Figure 2 shows $L_g$ for $X_{22}=0.02$ as a function of
$T_c$. The upper (lower) bound of the hatched region is for a pure
carbon (oxygen) WD and denotes the range of possibilities for
arbitrary C/O mixtures. The sharp downturn at $T_c\approx
(3-4)\times 10^{6}$K is due to core crystallization. We only allow for
sedimentation in the liquid parts of the WD, so that, as $T_c$
decreases and the solid core increases, $L_g$ decreases. If we
allowed for continuing sedimentation throughout the solidifying WD, 
the hatched
region would just continue to the left with the same slope. The dashed
line is the $L-T_c$ relation for a cooling $0.6M_\odot $ WD (see
caption). For $0.6M_\odot$, $L_g$ is always at least a factor of 10
lower than the exiting luminosity, $L$, at that $T_c$.

However, this competition gets closer as the WD mass increases. The
bottom panel of Figure 2 shows the same curves for $1.2M_\odot$. 
In this case, $L_g$ is a factor of 5 or so less than $L$, which
means that the WD cooling trajectory would be affected considerably.
It is a coincidence that $L_g$ drops off before becoming dominant and
then remains nearly parallel to the $L-T_c$ relation. This would not
occur in a pure helium WD, as crystallization never sets in. However,
in that case, there would be no \Ne present. It would just be the
settling of the other, less abundant, neutron rich metals, such as
$^{56}$Fe. Though it appears unlikely that the settling of iron 
will modify the cooling of a low-mass helium ($M<0.4M_\odot$) WD, 
substantial settling will occur in 10 Gyr. 

 The point we wish to make here is that the uncertain microphysics of
the diffusion coefficient clearly allows for an alternative view of
cooling for massive WD's. Namely, it is possible that the settling
luminosity can heat a white dwarf well enough to keep it in the liquid
state for a very long time. For example, increasing $D$ by a factor of
6 for the $1.2M_\odot$ WD would shift the $L_g$ curves up enough so
that the WD could be powered by sedimentation alone: hence, our
dubbing of these solutions as ``Sedimentars''. 
Or, if we allow for continued sedimentation once the core is
supposedly solid (say, it is supercooled), the lines would intersect
at $L\approx 2\times 10^{30} \ {\rm erg \ s^{-1}}$ for a $1.2M_\odot$
WD, which would then take a time in excess of 10 Gyr to radiate all of
the sedimentation energy. This sedimentar would live for many Gyr's 
at an effective temperature of 11,350 K. For a pure oxygen $1.36
M_\odot$ WD this crossover would occur at $L\approx 10^{31} {\rm erg \
s^{-1}}$ or $T_{\rm eff}\approx 21,000 {\rm K}$, still too cold to 
help explain the possible excess of hot, massive WD's reported by
Vennes (1999). 

\section{Conclusions}

 Our initial calculations suggest a number of important problems to
address in the future. The coincidence of the settling time with the
Hubble time forces us to reconsider the physics of the
inter-diffusion coefficients, especially in the quantum liquid
realm. This is critically important for the cooling of massive WD's
where there is some chance for the WD to shine for Gyrs on the power
of \Ne sedimentation. These gravitationally powered ``sedimentars''
would be a new astrophysical object that behave differently than the
conventional cooling WD.

  The time-dependent evolution of the \Ne abundance in different
astrophysical settings is the subject of our future work.  In that
paper, we evolve both cooling and accreting WD's with active \Ne
settling. This will allow for an accurate calculation of the evolution
of the \Ne abundance from starting conditions that reflect the prior
stellar evolution. For cooling WD's, we can follow the \Ne throughout
the star as it crystallizes, and assess the possibility of altering
the nature of the phase separation at freezing via enhanced \Ne
abundances.

The downward diffusion speed is very slow and it is natural to
question whether such slow speeds are actually relevant. We believe
so, as there is no expectation for active convection to undo
it. Comparably weak diffusive settling also appears in the Sun, where
inclusion of the slow gravitational settling of helium dramatically
improves the agreement of the solar interior model with that measured
by seismological inversions (Christensen-Dalsgaard et al. 1996; Basu,
Pinsonneault \& Bahcall 2000).

\acknowledgements
 For conversations about this problem, we thank
 N. Ashcroft, D. Chernoff, S.  Kulkarni, J. Liebert, E. Salpeter, and
 D. Stevenson. For discussions about current white-dwarf cooling
 models, we thank B.  Hansen and D. Townsley.  O. Straniero provided
 information on the \Ne distribution in the WD's and the anonymous
 referee's comments greatly improved our discussion. This research was
 partially supported by NASA via grant NAG 5-8658 and by the National
 Science Foundation under Grants PHY94-07194, PHY99-07949 and
 AY97-31632.  L. B. is a Cottrell Scholar of the Research Corporation.

\begin{figure}
\plotone{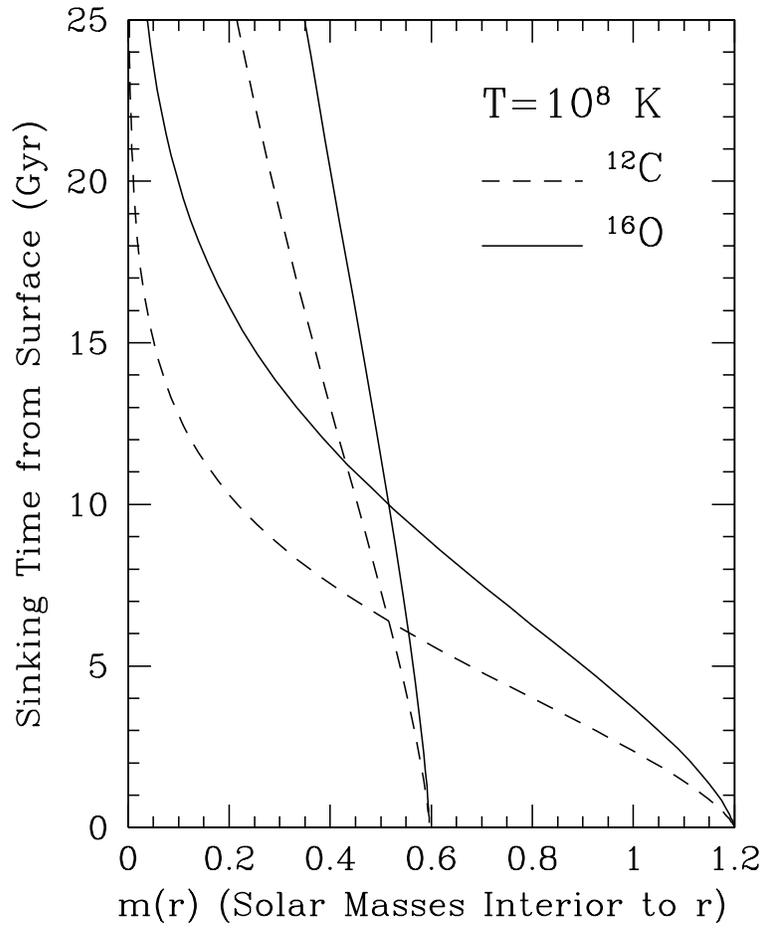}
\figcaption{The gravitational 
settling time for a \Ne nucleus in a pure C (dashed) or
pure O (solid) WD at a temperature of $10^8$ K appropriate for a Type
Ia progenitor. The lines show how
long it takes a \Ne nucleus to fall to a given mass coordinate if it
begins at the stellar surface. The set that begins at 1.2 (0.6)
$M_\odot$ is for a WD of that mass. 
\label{fig:time}}
\end{figure}

\begin{figure}
\plotone{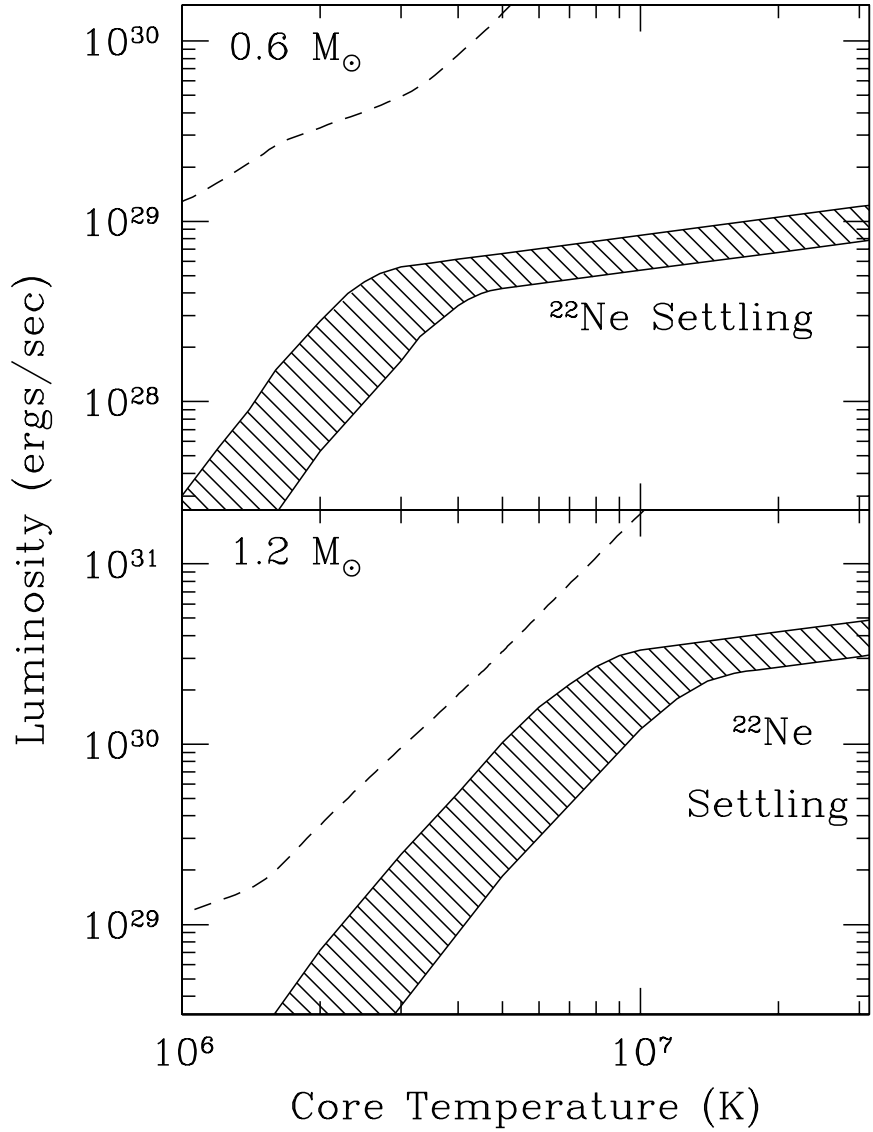}
\figcaption{The luminosity from \Ne settling in a $0.6M_\odot$ (top
panel) and $1.2M_\odot$ (bottom panel) WD with $X_{22}=0.02$.
The hatched region shows the 
luminosity from \Ne settling as a function of WD core temperature. The
upper (lower) bound is for a pure carbon (oxygen) WD and should 
bracket the range of possible C/O mixtures. The dashed lines are the
$T_c-L$ relations for cooling DA WD's from Benvenuto \& Althaus
(1999). Both have helium layers of $M_{\rm He}=10^{-2}M_\odot$, while the
$0.6(1.2)M_\odot$ WD has a hydrogen layer mass of $10^{-4}(10^{-6}) 
M_\odot$. The $0.6M_\odot$ model is close to that of
Hansen (1999) and Chabrier et al. (2000).
\label{fig:lum}}
\end{figure}

\end{document}